\documentclass[aps,prd,twocolumn,superscriptaddress,showpacs]{revtex4}

\usepackage{graphicx}
\usepackage{dcolumn}
\usepackage{bm}
\usepackage{natbib}

\newcommand{\be}{\begin{equation}}
\newcommand{\ee}{\end{equation}}
\newcommand{\bea}{\begin{eqnarray}}
\newcommand{\eea}{\end{eqnarray}}

\newcommand{\nhat}{\hat{\bf n}}

\newcommand{\edm}{\epsilon_{dm}}
\newcommand{\edmnow}{\epsilon_{dm,0}}
\newcommand{\sv}{\langle\sigma_Av\rangle}

\newcommand{\WMAP}{{\slshape WMAP~}}
\newcommand{\PLANCK}{{\slshape Planck~}}

\newcommand{\s}{{\rm ~s}}

\newcommand{\cm}{{\rm ~cm}}

\newcommand{\eV}{{\rm ~eV}}
\newcommand{\eVs}{{\rm ~eV/s}}
\newcommand{\keV}{{\rm ~keV}}
\newcommand{\MeV}{{\rm ~MeV}}
\newcommand{\GeV}{{\rm ~GeV}}
\newcommand{\TeV}{{\rm ~TeV}}

\newcommand{\sigmav}{\langle\sigma_Av\rangle}
\def\la{\vcenter{\hbox{$<$}\offinterlineskip\hbox{$\sim$}}}
\def\ga{\vcenter{\hbox{$>$}\offinterlineskip\hbox{$\sim$}}}

\begin{document}

\title{Detecting Dark Matter Annihilation with CMB Polarization : Signatures and
Experimental Prospects}

\author{Nikhil Padmanabhan}
\email{npadmana@princeton.edu}
\affiliation{Joseph Henry Laboratories, Jadwin Hall, Princeton University, 
Princeton, NJ 08544, USA}

\author{Douglas P. Finkbeiner}
\email{dfink@astro.princeton.edu}
\thanks{Henry Norris Russell Fellow, Cotsen Fellow}
\affiliation{Dept. of Astrophysical Sciences, Peyton Hall, Princeton University, 
Princeton, NJ 08544, USA}

\date{\today}

\begin{abstract}
Dark matter (DM) annihilation during hydrogen recombination ($z \sim 1000$)
will alter the recombination history of the Universe, and affect the 
observed CMB temperature and polarization fluctuations. Unlike other astrophysical probes of DM,
this is free of the significant uncertainties in modelling galactic physics, and
provides a method to detect and constrain the cosmological abundances of these particles. 
We parametrize the effect of
DM annihilation as an injection of ionizing energy at a rate $\epsilon_{dm}$,
and argue that this simple ``on the spot'' modification is a good approximation to the 
complicated interaction of the annihilation products with the photon-electron plasma. Generic models 
of DM do not change the redshift of recombination, but change the residual ionization
after recombination.
This broadens the surface of last scattering,
suppressing the temperature fluctuations and enhancing the polarization fluctuations.
We use the temperature and polarization angular power spectra to measure
these deviations from the standard recombination history, and therefore,
indirectly probe DM annihilation. The modifications to the temperature 
power spectrum are nearly degenerate with the primordial scalar spectral
index and amplitude; current CMB data are therefore unable to put any constraints on the 
annihilation power. This degeneracy is broken by polarization; \PLANCK will have the sensitivity
to measure annihilation power $\epsilon_{dm}(z=1000) > 10^{-15} \eV/{\rm s}/{\rm proton}$, while high sensitivity
experiments (eg. NASA's CMBPOL) could improve that constraint to $\epsilon_{dm}(z=1000) > 4\times 
10^{-16} \eV/{\rm s}/{\rm proton}$,
assuming a fractional detector sensitivity of $\Delta T/T \sim 1 \mu{\rm K}$ and a beam of 
$3'$. These limits translate into a lower bound on the mass of the DM particle, 
$M_{dm} > 10 ~-~ 100 \GeV$, assuming a single species with a cross section of
$\sv \sim 2 \times 10^{-26} {~\rm cm}^{3}/{\rm s}$, and a fraction $f \sim 0.1 ~-~1$ 
of the rest mass energy used for ionization. The bounds for the \WMAP 4y data are significantly lower,
because of its lack of high S/N polarization measurements, but it can
strongly constrain ${\cal O}(\MeV)$ particles such as those proposed by 
Boehm et al (2004).
\end{abstract}


\maketitle

\section{Introduction}

There is a broad consensus that the majority of the matter in the universe
is non-luminous, non-baryonic ``dark matter'' (DM) \citep[see][for recent reviews]{1996PhR...267..195J,
2004PhR...405..279B}. Some of the first compelling
evidence for DM came from galaxy rotation curves, suggesting that a large 
fraction of the mass lay beyond the luminous extent of the galaxy. Similar 
conclusions were reached for massive clusters of galaxies
using the gravitational distortion of the images of background
galaxies. Measurements of the
deuterium abundance, combined with big-bang nucleosynthesis indicate that 
density of baryons is less than the estimated mass density. This is supported by
measurements of the temperature fluctuations of the 
cosmic microwave background (CMB) radiation made by the \WMAP 
satellite\citep{2003ApJ...583....1B}. Dark matter is an integral part of the
cosmological ``standard model'', constituting 80\% of the total matter in the Universe.

Cosmology also constrains the gravitational properties of the DM.
The spatial clustering of galaxies and the angular power 
spectrum of CMB temperature fluctuations strongly favor a non-relativistic
pressureless species for the DM, while N-body simulations show that 
this clustering is consistent with the DM being composed of weakly interacting 
particles whose dominant long-range interactions are gravitational. This
suggests that the DM particle is a weakly interacting
massive particle (WIMP), although more exotic possibilities are not
ruled out. We, however, will focus solely on WIMPs for this paper.

A theoretical description of the DM is still unknown, although it is 
believed that such a description involves physics beyond the 
Standard Model. There are strong theoretical reasons for believing that the Standard Model
is modified at the electroweak symmetry scale $\sim 1 \TeV$, intriguingly
the same energy scale for WIMPs predicted by the present mass density.
These modifications range from supersymmetric extensions of
the Standard Model to large extra dimensions modifying gravity on these scales,
and generically have a zoo of 
massive particles with properties that make them potential 
DM candidates. The relevant energies are just entering 
the reach of current direct detection experiments\citep{2004PhRvL..93u1301A}
and accelerator energies, and will be strongly constrained by
the next generation of these experiments.

While there is no substitute for a direct detection of WIMPs, astrophysical
probes of DM play an important
role, since they provide a complementary view of the DM parameter space, 
making different assumptions than particle physics experiments.
Furthermore, probing the cosmological abundance of
a candidate particle  requires an astrophysical 
probe. There already are a few tantalizing observations
suggesting a DM particle between $\sim 1 ~-~ 100 \GeV$. The $\gamma$-ray
emission measured by the EGRET instrument on the \emph{Compton Gamma
Ray Observatory} has a higher amplitude at $\ga 1\GeV$ than
traditional models of the Galactic cosmic ray population and
interstellar medium models allow \citep{1997ApJ...481..205H, 2000ApJ...537..763S,
2004astro.ph.12620D}.
The $e^\pm$ annihilation
line strength at $511\keV$ observed by Integral/SPI suggests a
surprising Galaxy-wide positron production rate of $10^{44}\s^{-1}$
\citep{2004astro.ph.10354T}.  Furthermore, the cosmic-ray positron ratio
observed by HEAT shows an excess above
$5\GeV$ consistent with simple models of Higgsino decay \citep{2004PhRvL..93x1102B}.
The synchrotron haze in the inner Milky Way \citep{2004ApJ...614..186F} may be an example
of synchrotron emission from $e^\pm$ pairs produced by ongoing DM annihilations 
as suggested by \citep{2004PhRvD..70b3512B}.
The observed synchrotron signal could
be produced by a fiducial model of $100\GeV$ particles 
annihilating at $\sigmav=2\times 10^{-26}\cm^3\s^{-1}$ and distributed with
an NFW\citep{1997ApJ...490..493N} mass profile\citep{2004astro.ph..9027F}. This model is not expected
to be correct in detail, but is a fiducial model scalable
to other possible scenarios.

In this paper, we propose using the CMB temperature and polarization fluctuations 
as a probe of DM annihilation. 
DM annihilation at $z \sim 1000$ injects energy into the photon-baryon plasma ionizing
neutral hydrogen and modifies the recombination history.
These additional electrons scatter CMB photons, 
making the last scattering surface thicker and attenuating correlations between
temperature perturbations. On the other hand, the correlations 
between polarization fluctuations are enhanced by the thicker scattering surface. This alters
the temperature and polarization angular power spectra, 
providing a handle on the properties of the DM. The 
CMB  has an important advantage over the other probes discussed above, as
computing the power spectrum is a linear calculation based
on well understood physics. This enables detecting small deviations
from the expected signal (that herald new physics) at a high level 
of significance. This should be contrasted with the other probes discussed above that
require additional information such as the DM distribution and clumpiness, ISM density, magnetic field 
strength and degree of tangling, Galactic photon energy density etc.,
all of which are complex processes with significant uncertainties.

Modifications to the recombination history, ranging from
delayed recombination to low redshift ionization from DM decays, and their effect of the CMB have 
been considered by \cite[eg.][]{2000ApJ...539L...1P,2003PhRvD..68h3501B, 2004ApJ...600...26H,
2004PhRvL..92c1301P, 2004PhRvD..70d3502C,2003ApJ...586..709D}. Most recently, these
ideas have been used to attempt to explain the high optical depth observed by the \WMAP satellite
by using DM decays to reionize the universe. The goals of this paper
significantly differ from such studies. We aim to understand how the 
CMB (via the recombination history) constrains the parameters of an annihilating DM model. 
Since we lack a preferred theoretical framework for the DM,
we construct a generic parametrization of DM annihilation
to effectively constrain the DM model
space. We parametrize DM annihilation as an energy injection
into the IGM\footnote{The use of
``Inter-Galactic Medium'' for the photon-baryon fluid at $z\sim 1000$ is
a convenient anachronism; we use it throughout without further apology.}
and relate this energy to standard
parameters like the DM mass and annihilation cross section.
Given a specific model of the DM, constraints on this injection energy
can then be used to constrain the parameters of that model.

We start (Sec.~\ref{sec:ionize}) by discussing
the effect of DM annihilation on the ionization history of the Universe, 
introducing the ``on the spot'' approximation as a useful parametrization
of the more complex processes responsible.
Sec.~\ref{sec:CMB} then discusses the effect of this change to the 
ionization history on the CMB, while Sec.~\ref{sec:detect} attempts 
to quantify the detectability of these effects. Sec.~\ref{sec:other} 
considers whether other astrophysical probes have the potential to probe
DM annihilation during the recombination epoch, and Sec.~\ref{sec:discuss}
summarizes the principal conclusions of this paper, and discusses their
implications.

A note on cosmology and notation: if not explicitly specified, we will assume a
concordance fiducial model with $\Omega_{M} + \Omega_{\Lambda} =1$,
$\Omega_{M}=0.3$, $\Omega_{b} = 0.05$, $h=0.7$, $Y_{He} = 0.24$ and
$n_{s}=1$. We only consider the scalar contributions to the 
CMB fluctuations, and therefore
describe the temperature and polarization fluctuations by 
the temperature-temperature (TT), temperature-polarization (TE), and 
polarization-polarization (EE) power spectra. Finally, proper and comoving
times are $t$ and $\eta$ respectively, while a $0$ subscript denotes the
present epoch.

\section{The Ionization History}
\label{sec:ionize}

The effect of DM annihilation on the recombination history can 
be conveniently, albeit artificially,  separated into two stages -- the 
injection of the energy from the annihilation into the IGM, 
and the effect of this energy on recombination. We argue that the former process
is well approximated by a rate of energy injection per hydrogen atom ($\edm$),
instantaneously used to ionize and heat the IGM. We then consider how this energy
changes the ionization fraction as a function of time.

\subsection{The ``On the Spot'' Approximation}

\begin{figure}
\includegraphics[width=3in]{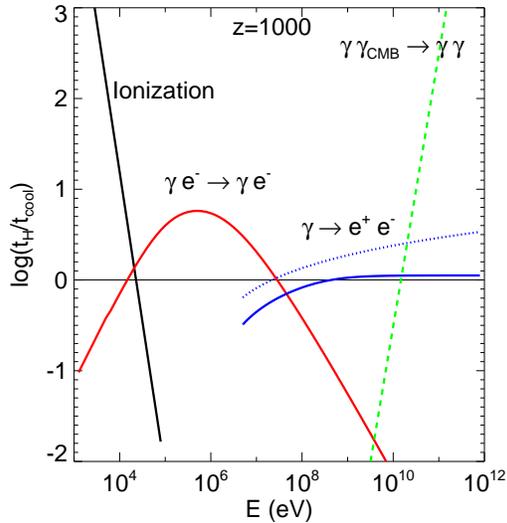}
\caption{\label{fig:photonlosses} A comparison of the photon cooling time to the Hubble time at $z=1000$,
for different photon energies. The dominant processes (in order of increasing energy) are ionization, 
Compton scattering, pair production, and photon-photon scattering. All the curves (except the 
dotted curve) assume a neutral IGM,
with a  density of $2 \times 10^{-7}$ cm$^{-3}$ atoms today. The dotted curve shows the
pair production rate for a completely ionized IGM. Regions where $t_{H}/t_{cool} < 1$ are
transparent; photons injected at these energies lose their energy by redshifting.
Note that photon-photon scattering does
not transfer the energy to electrons, but simply redistributes it to lower energies. This figure
ignores pair production off CMB photons since this process is subdominant for the energy range
considered here; it however dominates at higher energies.}
\end{figure}

How do DM annihilations cause ionizations? The primary products from an annihilation depend on
the particular DM model, but generically are quarks,
gauge bosons, leptons, and Higgs particles. These primaries tend to be unstable, and
rapidly decay via hadronic-leptonic jets into showers of $e^{\pm}$ pairs, protons, photons and 
neutrinos. Given our ignorance about DM, we assume
each annihilation partitions the majority of its energy between $e^{\pm}$ pairs, photons and 
neutrinos, whose energy spectra can be calculated given the mass
and couplings of the DM particle. Note that we are not assuming that these are directly 
produced by the annihilation, but simply that they are final products of the 
resulting particle cascades. The problem now simplifies to 
understanding the mechanisms by which $e^{\pm}$ pairs, photons and neutrinos inject energy into 
the IGM. Of these, neutrinos are the easiest to understand; they 
never interact and their energy is lost.

The interaction of photons with the IGM was considered in detail by \cite{1989ApJ...344..551Z}
who find that the dominant processes (ordered by increasing photon energy) are photoionization, 
Compton scattering, pair production off nuclei and atoms, photon-photon scattering, and pair
production off CMB photons (Fig.~\ref{fig:photonlosses}). To estimate the efficiency of these mechanisms, we compare the
cooling time for each process, $t_{cool} \equiv 1/(d \ln E/dt)$, to the Hubble time, $t_{H} \equiv 1/H(z)$.
Except for Compton scattering, we approximate the cooling time by the mean free time
as most of the energy is lost in the first interaction.
If $t_{H} \gg t_{cool}$ (Fig.~\ref{fig:photonlosses}), energy 
deposition is very efficient, either by directly ionizing the IGM, or by producing energetic electrons.
Conversely, if $t_{H} \ll t_{cool}$, the universe is optically thin and most of the energy is lost 
through the redshifting of photons and not to ionizations. 
The photon-photon scattering process \citep{1990ApJ...349..415S}
is an exception to the above - 
each scattering event, on average,  equally divides the energy between the two photons. 
The photon energy spectrum therefore gets shifted to lower energies until either pair production starts 
to dominate, or the universe becomes transparent.

What happens to photons injected into the transparency window 
between $\sim 10^{8} ~-~ 10^{10} \GeV$ (Fig.~\ref{fig:photonlosses})? The ratio $t_{H}/t_{cool} 
\propto (1+z)^{3/2}$ ($\propto (1+z)^{9/2}$ for two photon scattering), while the photon
energy redshifts as $(1+z)$. These photons therefore remain in the optically thin regime, 
and contribute to the diffuse photon background today (Sec.~\ref{sec:other}).

The second component of energy injection comes from electrons
\footnote{Positrons behave identically to electrons at high energies, while 
at low energies, they annihilate releasing $2\times 511 \keV$ in photons.}, 
both from the 
annihilation products, as well as from Compton scattering and pair production considered
above. Their energy loss has been considered by a number of authors
\cite{1970RvMP...42..237B, 1985ApJ...298..268S, 2004PhRvD..70d3502C}; we 
restrict ourselves to a brief discussion of the relevant processes and time scales. 
At high electron energies ($\gamma \gg 1$), the dominant energy loss is 
by inverse Compton scattering CMB photons. The cooling time is \cite{1970RvMP...42..237B},
\be
\left(\frac{1}{t_{cool}}\right) = \frac{- d \ln \gamma}{dt} = \frac{4 \sigma_{T} c a_{R} T_{CMB}^{4} 
\gamma}{3 m_{e} c^{2}} \,\,,
\ee
where $T_{CMB} = 2.725 (1+z) {\rm K}$ is the mean CMB temperature at the relevant redshift, $a_{R}$
is the radiation constant, and $\sigma_{T} = 6.65\times 10^{-25} \cm^{2}$ is the
Thomson cross section. 
Comparing this to the Hubble time, one finds
\be
\frac{t_{H}}{t_{cool}} \sim 10^{5} \left( \frac{1+z}{1000} \right)^{5/2}
\frac{1}{\sqrt{\Omega_{M}h^{2}}} \gamma \,\,\,,
\ee
implying that inverse Compton cooling efficiently produces
photons with energies
\be
E_{\gamma} \sim 5 \left(\frac{1+z}{1000}\right) \left(\frac{E_{e}}{1 \GeV}\right)^{2} {\rm MeV} \,\,.
\ee
Fig.~\ref{fig:photonlosses} shows that electrons with energies $< 100\MeV$ will produce
photons that efficiently ionize hydrogen; above that energy, the scattered photons either produce
an electromagnetic cascade by Compton scattering or pair production, or are scattered into the 
optically thin part of the spectrum from $10^{8}$ to $10^{10}\eV$ and escape.

At lower energies, the principal mechanisms for energy loss become collisional heating,
excitations and ionizations\cite{1985ApJ...298..268S}.
At high kinetic energies, $E_{e} \gg 100 \eV$, the cross section
for collisional ionization is \cite{1985ApJ...298..268S},
\be
\sigma_{eH} = \frac{2.23 \times 10^{-15} \ln(E/13.6)}{E} {\rm cm}^2\,\,,
\ee
($E$ measured in ${\rm eV}$) implying $t_{H}/t_{cool} \gg 1$ at $z \sim 1000$.
The results for collisions and
excitations are very similar, although collisional losses become important
at lower energies, $100\eV < E < 1\keV$.

\begin{figure}
\includegraphics[width=3in]{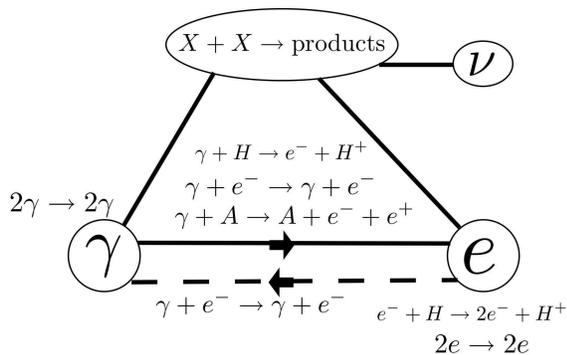}
\caption{\label{fig:cascade} The injection of energy from dark matter annihilation 
into the IGM, via the creation of electromagnetic cascades. Energy transfer to the IGM takes place 
principally through the ionization and collisional processes.} 
\end{figure}

Our model of DM annihilation, summarized in Fig.~\ref{fig:cascade}, is
\begin{enumerate}
\item DM particles annihilate into jets, whose
end products are dominated by electrons, photons and neutrinos.
\item The IGM is transparent to neutrinos; this energy is lost.
\item The photons and electrons trigger electromagnetic cascades, shifting 
their spectra to lower energies, until their energy is either deposited in 
the IGM, or is redshifted away when the photons enter an optically thin regime.
\item The time scales for the cascades and energy deposition are much 
smaller than the expansion time.
\end{enumerate}
We parametrize the effect of DM annihilation by the
rate of energy injection per hydrogen nucleus per time, $\edm$; furthermore, we
assume this energy is instantaneously deposited into the IGM. This
``on the spot'' approximation has 
the virtue of being generic and independent of particular properties of
the DM. 

What is the magnitude and redshift dependence of $\edm$? Given a particle
with mass $M_{dm}$ and a thermally averaged cross-section $\sv$, we obtain,
\be
\edm = f M_{dm} \left( \frac{\sv n_{dm,0}^2}{n_{H,0}} \right) (1+z)^3 \,\,,
\label{eq:dmtoedm}
\ee
where $n_{DM,0}$ and $n_{H,0}$ are the present densities of the dark matter and hydrogen
particles respectively, and $f$ is the fraction of the rest mass energy injected into 
the IGM. Assuming our fiducial cosmology with a single DM species, we find,
\bea
\edm  \sim  f 10^{-24} (1+z)^{3}\,\eV\s^{-1} \nonumber \\
\times \left[ \left(\frac{100\GeV}{M_{dm}}\right) 
\left(\frac{\sv}{2 \times 10^{-26} \cm^3\s^{-1}} \right) \right] \,\,.
\label{eq:dmtoedm2}
\eea
The injected energy is inversely proportional to the particle mass; more massive
particles inject \emph{less} energy into the IGM.
We parametrize our ignorance of the annihilations and their effect on the
IGM by a simple efficiency factor, $f$. Given a 
specific model, one can compute $f$ and convert constraints on $\edm$ into 
constraints on $M_{dm}$ and other model parameters.

\subsection{Recombination with DM annihilation}

Given $\edm$, we compute its effect on the recombination history. This energy injection
heats the IGM, and ionizes and excites the hydrogen and helium atoms.
\cite{1985ApJ...298..268S} compute the exact fractions converted to heat, 
ionization and excitation as a function of the ionization fraction, and find 
that for a neutral IGM, the energy is roughly equipartitioned between the three 
processes, while for a fully ionized plasma, all the energy is converted into heat.
This suggests a simpler but adequate approximation \cite{2004PhRvD..70d3502C}
that $(1-x)/3$ of the energy goes into ionization and $(1+2x)/3$ into
heating the IGM, where $x$ is the ionization fraction and we assume that excitations neither
change the matter temperature nor the ionization fraction.

\begin{figure}
\includegraphics[width=3in]{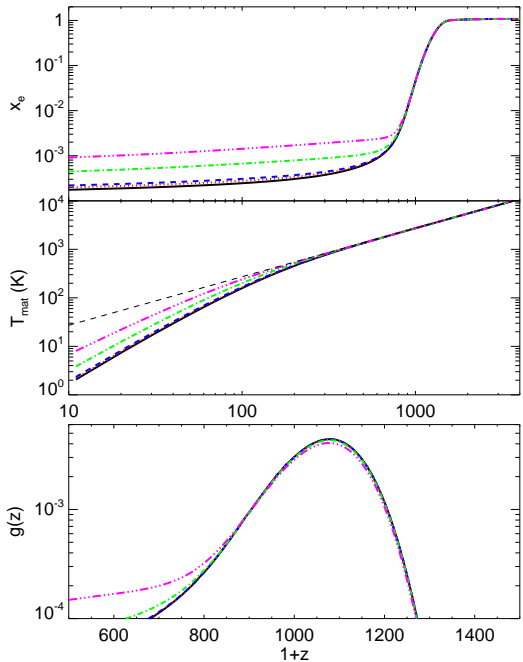}
\caption{\label{fig:recvis2} The ionization fraction $x_{e}$ (top), matter temperature (center), 
and visibility function (bottom) as a function of $\edm$. The heavy solid lines show the fiducial
model with $\edm=0$; from bottom to top, $\edmnow = 5, 10, 100, 500 \times
10^{-25}$ eV/s. The thin dashed line in the center plot shows the evolution of CMB temperature, $T(z) = 
T_{0}(1+z)$. Note that the injection of additional energy does not slow recombination, but increases the
residual ionization; this leaves the peak of the visibility function unchanged but broadens the surface of
last scattering.}
\end{figure}

We compute the recombination history using the public code \texttt{RECFAST} 
\cite{1999ApJ...523L...1S}
modifying the evolution equations as follows,
\bea
-\delta \left(\frac{dx[H]}{dz}\right) = & & \nonumber \\
& &\frac{\edmnow}{13.6} \frac{1-x[H]}{3(1+f_{He})} {\cal F}(z)\,,  \nonumber \\ 
-\delta \left(\frac{dx[He]}{dz}\right) = & & \nonumber \\
& &\frac{\edmnow}{24.6} \frac{1-x[He]}{3(1+f_{He})} {\cal F}(z) 
\eea
where  $\edmnow$ is the energy injection rate at the present epoch (in eV/s), 
and 
\be
{\cal F}(z) \equiv \frac{(1+z)^{3}}{H(z)(1+z)} \,\,.
\ee
Additionally, the ionization fraction of a species $A$ is defined as,
\be
x[A] = \frac{n[A^{+}]}{n[A^{+}] + n[A]} \,\,,
\ee
and $f_{He}$ is the ratio of the 
number density of helium to that of hydrogen. The evolution of the 
matter temperature, $T_{m}$ is similarly given by,
\bea
& & -\delta \left(\frac{dT_{m}}{dz}\right) = \nonumber \\
& &\frac{2 \edmnow}{3 k_{B}} \frac{1+2x[H]+f_{He}(1+2x[He])}{3(1+f_{He})}{\cal F}(z) \, .
\eea

The resulting recombination and matter temperature histories for different values of $\edmnow$ are
shown in Fig.~\ref{fig:recvis2}; the dominant effect
is to change the residual ionization after recombination. This is 
easily explained by considering the competition between the recombination 
rate and the expansion of the universe. At early times, the recombination rate is significantly
greater than the expansion rate and therefore, additional ionizations due to 
DM annihilation are immediately erased. As the recombination rate slows, these additional
ionizations ``freeze out'', leading to a greater residual ionization fraction.

The evolution of the matter temperature is similar. At redshifts $\gg 100$, Compton scattering
keeps the matter and radiation in tight thermal contact, and the 
excess energy from DM annihilation is lost in the extremely large heat capacity of the
blackbody radiation. However, as the matter completely decouples from the radiation, 
annihilations start to increase the matter temperature, resulting 
in slower cooling relative to the fiducial model.

\section{The CMB as a probe}
\label{sec:CMB}

Having computed the effect of DM annihilation on the recombination history, we attempt
to understand its effect on the CMB. In what follows, it is sometimes convenient to
parametrize the effect of DM annihilation by an
ionization ``floor'' added to the standard recombination history. 
This lacks the physical intuition of $\edm$, but is a convenient analytic approximation.
We define the optical depth to Thomson scattering,
\be
\tau(\eta) = \int_{\eta}^{\eta_{0}} \, d\eta \, \sigma_{T} n_{e} c a \,\,,
\ee
where $n_{e}$ is the free electron density. Assuming a matter dominated cosmology
and constant ionization fraction $x_{e}$, this gives us
\be
\tau(z) \sim 4 \times 10^{-2} x_{e} \frac{\Omega_{b} h (1-Y_{He})}{\sqrt{\Omega_{M}}}
z^{3/2} \,\,,
\label{eq:nicetau}
\ee
if $z \gg 1$.

\subsection{Peak Positions}
\label{sec:peak}

We begin by estimating the change in the position of the acoustic peaks in the
temperature power spectrum due to an ionization floor.
The probability that a photon last scattered between redshifts $z$ and 
$z+dz$ is given by the visibility function,
\be
g(z) \equiv \tau'(z) e^{-\tau(z)} \,\,,
\label{eq:vis}
\ee
shown in Fig.~\ref{fig:recvis2} for different recombination histories.
The fraction of photons that scatter at a redshift $<z$, $G(z)$,
is simply the integral of visibility function, $G(z) = 1 - \exp(-\tau(z))$. Since
$g(z)$ is sharply peaked, we can meaningfully define a redshift
of last scattering, $z_{LS}$, that determines the angular positions of the acoustic
peaks. A convenient definition is $G(z_{LS})=0.5$ or
$\tau(z_{LS}) \approx 0.7$ implying $z_{LS} \sim 1050$ for standard recombination.
For the ionization floor to significantly shift the peaks, the
additional optical depth, $\Delta \tau$, would have to be $\sim 1$. 
Using Eq.~\ref{eq:nicetau}, this requires
$10^{3} x_{e,floor} \sim (\sqrt{\Omega_{M}}/\Omega_{b} h)$, or $x_{e,floor}
\sim 0.01$ for our fiducial cosmology.
As we shall see below, such an ionization fraction would have already noticeably
affected the CMB temperature and polarization
and therefore is strongly disfavored. More plausible
values of the ionization floor do not noticeably shift the positions of the acoustic
peaks in the temperature power spectrum.

\begin{figure}
\includegraphics[width=3in]{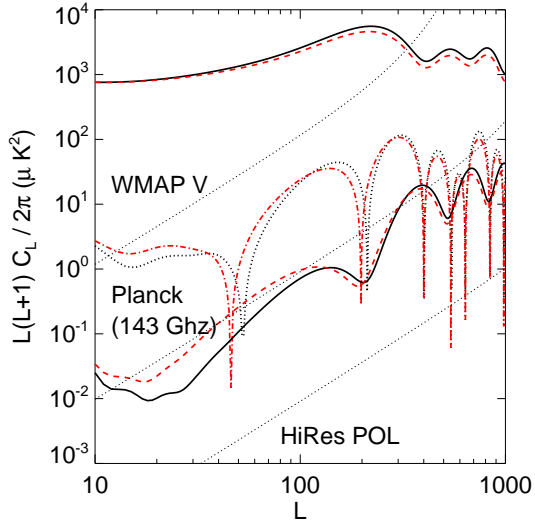}
\caption{\label{fig:models} The TT, TE, and EE angular power spectra for our fiducial 
cosmological model, with no DM annihilation (solid and dotted lines), and with 
$\edmnow=10^{-22} \eVs$. Also shown are the polarization noise spectra, for 
the \WMAP V band, the \PLANCK 143 Ghz channel, and a hypothetical high resolution
polarization experiment (see Table \ref{tab:expspec} for details).}
\end{figure}

\subsection{Power Spectra}

The effect of the altered recombination history on the 
CMB power spectra is discussed analytically below.
However, the numerical results presented in the paper use the publicly available
Boltzmann code \texttt{CAMB} \cite{2000ApJ...538..473L}, with the modified version of \texttt{RECFAST}
described in the previous section, to obtain accurate power spectra. An example
is shown in Fig.~\ref{fig:models}.

The temperature angular power spectrum is the photon distribution
function convolved with the visibility function (the last scattering 
surface), and projected on the sky. The photon distribution function is unchanged
by DM annihilation, but the visibility function extends to lower redshifts, broadening
the surface of last scattering. This suppresses perturbations
on scales smaller than the width of the surface, resulting in a relative
attenuation of the power spectrum.
This is scale dependent, with the largest scales attenuated the 
least and small scales the most. These effects are clearly seen in the accurate
numerical solutions in Fig.~\ref{fig:models}.

Given the imminent high S/N temperature measurements due from the \WMAP and \PLANCK
satellites, an immediate question is whether DM annihilation is detectable just using the
temperature power spectrum. Unfortunately, the 
effects of $\edm$ described above are almost perfectly degenerate with the 
slope and amplitude of the primordial power spectrum. To see this quantitatively,
we start with the line of sight solution
to the temperature perturbation in direction $\nhat$ \cite{1994ApJ...435L..87S},
\be
\Delta(\nhat) = \int^{\eta_{0}}_{0} \left[ \dot{\tau} \left(\Psi +
\frac{\Theta}{4} + \nhat \cdot {\bf v}_{b}\right) + 2\dot{\phi}
\right] e^{-\tau} \, d\eta \,\,,
\ee
where $\Psi$ is the gravitational potential, $\Theta$ is the photon
density perturbation, ${\bf v}_{b}$ is the baryon velocity, and we
ignore vector and tensor contributions. If we ignore the ISW \cite{1967ApJ...147...73S}
contribution ($2 \dot{\phi}$) to the anisotropy spectrum, we obtain a
useful semi-analytic approximation to the anisotropy spectrum by
separating into slowly varying (potentials, $T(k)$ below) and
rapidly varying (recombination, Silk damping, $D(k)$ below)
terms \cite{1994ApJ...435L..87S},
\be
C_{l} = 4\pi A \int^{\infty}_{0} \, d({\rm ln}\, k) \, k^{n_{s}} D^{2}(k)
T^{2}(k) \,\,,
\label{eq:clanalytic}
\ee
implicitly assuming that $T^{2}(k)$ is evaluated at the redshift
of last scattering and has no time dependence. The damping function is
given by \cite{1995ApJ...444..489H},
\be
D(k) = \int\,dz\, g(z) \exp\left(-\frac{k^{2}}{k_{D}^2(z)}\right) \,\,,
\label{eq:dampk}
\ee
where $g(z)$ is the visibility function introduced earlier, and $k_{D}$
is the Silk damping scale given by \cite{1995PhRvD..52.3276Z},
\be
\frac{1}{k_{D}^{2}} = \int_{z}^{\infty} \,dz\, \frac{c}{H^{2}(z)} \frac{1}{6(1+R)\tau'(z)}
\left[\frac{R^{2}}{(1+R)} + \frac{16}{15}\right] \,\,,
\label{eq:silkdamp}
\ee
where $R = 3 \rho_{b}/4 \rho_{\gamma}$ is the baryon-photon ratio. Since the ionization
history only appears in Eq.~\ref{eq:clanalytic} through the optical depth in $D(k)$, we
estimate the effect of adding an ionization floor by computing $D(k)/D_{0}(k)$,
where $D_{0}(k)$ assumes the standard ionization history.

\begin{figure}
\includegraphics[width=3in]{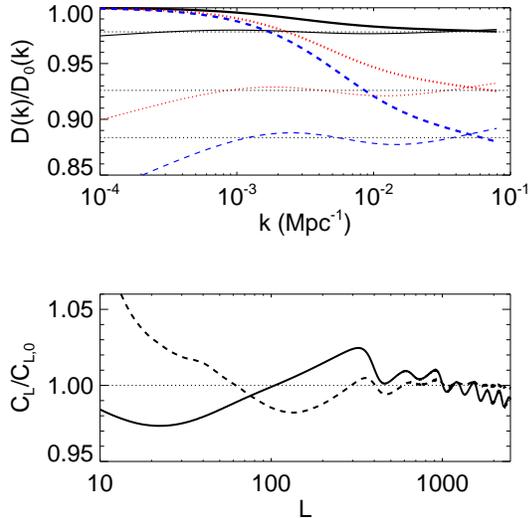}
\caption{\label{fig:dampfn} (Top) The heavy lines show the
ratio of damping functions $D(k)/D_{0}(k)$ for $\edmnow$ of $100$, 
(solid), $500$ (dotted), and $1000 \times 10^{-25} {\rm eV/s}$ (dashed) and our
fiducial cosmology. The light lines
show this same ratio divided by $(k/k_{fid})^{-\alpha}$ where $k_{fid} = 0.05$ Mpc$^{-1}$.
The horizontal dotted lines are a visual guide. (Bottom) The solid line shows the ratio of a
model with no DM annihilation, but with $n_{s}$ altered using the analytic calculation above, 
to our fiducial model with $\edmnow=500 \times 10^{-25} {\rm eV/s}$. The dashed line has the 
same ratio, except that all the cosmological parameters are adjusted to best fit the model with
DM annihilation.}
\end{figure}

As the relevant regime is when the ionization fraction is rapidly changing, we
numerically integrate Eq.~\ref{eq:dampk} and compute $D(k)/D_{0}(k)$
for different $\edmnow$. The results for our fiducial cosmology
are shown in Fig.~\ref{fig:dampfn}. The scales relevant for $l > 50$ in the CMB correspond
approximately to $k > 0.001 h {\rm Mpc}^{-1}$; 
Fig.~\ref{fig:dampfn} demonstrates that $D(k)/D_{0}(k)$ is
remarkably well described by a power law, $k^{-\alpha}$, over these scales. This signals 
a near exact degeneracy in the CMB; examining Eq.~\ref{eq:clanalytic}
suggests that the effect of the ionization floor can be almost exactly compensated
by adjusting $n_{s} \rightarrow n_{s} + 2\alpha$, and changing the amplitude, $A$. The residual
differences can be corrected by adjusting (at sub-percent levels) the remaining cosmological
parameters.

We emphasize that this degeneracy appears to be purely accidental. As $k\rightarrow 0$,
Silk damping becomes increasingly unimportant and $D(k)/D_{0}(k) \rightarrow 1$. In addition,
we have ignored the ISW contribution, which has a different visibility function, and
therefore will not be compensated by changing the scalar spectral index. On small scales ($k
\rightarrow \infty$) that are considerably damped before recombination, the correction
to the visibility function due to the ionization floor is negligible and again, one
would expect $D(k)/D_{0}(k) \sim$ constant. These two limits are however hard to constrain,
large scales because of cosmic variance, and small scales because of secondary anisotropies.
On intermediate scales where high S/N measurements of CMB can be made, the effect of an
ionization floor is degenerate with changing the scalar spectral index.

Estimating the effect of $\edm$ on the polarization of the CMB is more involved. 
Polarization principally results from the Thomson scattering of the local 
quadrupole in the temperature distribution. 
However, the quadrupole vanishes during the tightly coupled regime 
before recombination; the only source of a 
quadrupole is the free streaming of the monopole and dipole perturbations 
during recombination. Ignoring the effects of reionization, the amplitude of the 
quadrupole contributing to polarization can be schematically written as
\cite{2003moco.book.....D},
\be
\Theta_{2}(k) \sim \Theta_{0}(k) j_{2}(x) + 3 \Theta_{1}(k)\left[
j_{1}(x) - \frac{3j_{2}(x)}{x} \right] \,\,
\ee
where $\Theta_{l}$ with $l=0,1,2$ represents the monopole, dipole
and quadrupole components of photon distribution, and $x = k \Delta \eta$,
where $\Delta \eta$ is the thickness of the last scattering surface. 
Focusing on scales much larger than the thickness of the last scattering
surface, $x \ll 1$, we obtain,
\be
\Theta_{2}(k) \sim \frac{\Theta_{0}(k) [k \Delta \eta]^{2} + 
6 \Theta_{1} [k \Delta \eta]}{15} + {\cal O}(x^{3}) \,\,,
\label{eq:nicepol}
\ee
where we used the expansion $j_{l}(x) = x^{l}/(2l+1)!! + {\cal O}(x^{l+2})$.
Increasing the width of the last
scattering surface therefore increases the amplitude of the polarization 
fluctuations. Furthermore, Eq.~\ref{eq:nicepol} implies that
the quadrupole is dominated by free-streaming from the dipole perturbations.
These are $\pi/2$ out of phase with the monopole, resulting in 
the well known phase structure of the CMB temperature and polarization spectrum
peaks. As the last scattering surface grows thicker, the fractional contribution from
monopole perturbations to the quadrupole increases, shifting the positions of the TE and 
EE peaks. Finally, on smaller scales, the TE and EE power spectra are attenuated by 
increased scattering, analogous to the TT power spectrum. These trends
are seen in the TE and EE power spectra in Fig.~\ref{fig:models}.

\section{Estimating Detectability}
\label{sec:detect}

\subsection{Formalism}

Given a measurement of the
CMB sky with detector noise and cosmic variance, how distinguishable are any two models?
And does there exist a combination of standard CMB parameters that can mimic 
DM annihilation?
To answer the first question, we assume a realization of the full 
sky both in temperature and polarization (E modes). The likelihood of 
observing these maps is,
\be
{\cal L}({\bf d} | {\rm theory}) \propto \frac{1}{\sqrt{{\rm det}~ {\bf C}}} 
\exp\left(-\frac{1}{2} {\bf d}^{t} {\bf C}^{-1} {\bf d} \right) \,\,,
\ee
where we have assembled the maps into the vector ${\bf d}$, 
and the covariance matrix, ${\bf C}$, is a function of the 
theoretical model.
We transform to the spherical harmonic basis, 
\be
{\cal L}({\bf d}| {\rm theory}) \propto \prod_{l,m} \frac{1}{\sqrt{{\rm det}~{\bf C}_{l}}} 
\exp\left(-\frac{1}{2} {\bf d}^{t}_{lm} {\bf C}_{l}^{-1} {\bf d}_{lm} \right) \,\,,
\ee
where ${\bf d}^{t}_{lm} = (a^{T}_{lm}, a^{E}_{lm})$ is the spherical transform of the 
temperature and E-mode maps respectively. The $2 \times 2$ covariance matrix 
for each $l,m$ mode is given by
\be
{\bf C}_{l} = \left(
\begin{array}{cc}
C^{TT}_{l} & C^{TE}_{l} \\
C^{TE}_{l} & C^{EE}_{l} 
\end{array} \right) \,\,,
\ee
where $C^{TT}$, $C^{EE}$, and $C^{TE}$ are the temperature and E-mode angular
auto-power spectra and cross-power spectrum respectively. Taking logarithms 
and summing over azimuthal modes, we get
\bea
\log[{\cal L}(d | C)] = -\frac{1}{2} \sum_{l} (2l+1)  
\left[ \vphantom{\frac{1}{2}} \log({\rm det}~{\bf C}) + \nonumber \right. \\
\left. \frac{C^{TT}\hat{C}^{EE} + C^{EE}\hat{C}^{TT} - 2 C^{TE}\hat{C}^{TE}}{{\rm det}~{\bf C}} \right] \,\,,
\eea
where ${\rm det}~{\bf C} = C^{TT}C^{EE} - (C^{TE})^{2}$, the hat denotes observed quantities,
and the $l$ dependence is suppressed. 

We can now compare the likelihoods for two different theoretical models, ${\bf C}$ and ${\bf C}'$,
\be
r \equiv \log \left( \frac{{\cal L}(d | {\bf C})}{{\cal L}(d | {\bf C}')} \right) \,\,,
\ee
obtaining 
\bea
r = -\frac{1}{2} \sum_{l} (2l+1) \left[ 
\log\left(\frac{{\rm det}~{\bf C}}{{\rm det}~{\bf C}'}\right) + \right.\nonumber \\
\left. \alpha \hat{C}^{TT} + \beta \hat{C}^{EE} + \gamma \hat{C}^{TE} \vphantom{\frac{1}{2}} 
\right] \,\,,
\eea
where
\bea
\alpha = \left(\frac{C^{EE}}{{\rm det}~{\bf C}} - \frac{C^{EE'}}{{\rm det}~{\bf C}'}\right) \,\,, \\
\beta = \left(\frac{C^{TT}}{{\rm det}~{\bf C}} - \frac{C^{TT'}}{{\rm det}~{\bf C}'}\right) \,\,, \\
\gamma = -2 \left(\frac{C^{TE}}{{\rm det}~{\bf C}} - \frac{C^{TE'}}{{\rm det}~{\bf C}'} \right) \,\,.
\eea
If we assume that the data is a realization of ${\bf C}$, we have a simple criterion
for distinguishability (and therefore, detectability) -- two models are distinguishable
if the probability that $r< 0$, $P(r<0)$, is less than a chosen threshold (the confidence 
level). To compute $P(r<0)$, we compute 
\bea
\langle r \rangle = -\frac{1}{2} \sum_{l} (2l+1) \left[
\log\left(\frac{{\rm det}~{\bf C}}{{\rm det}~{\bf C}'}\right) + \right.\nonumber \\
\left. \alpha C^{TT} + \beta C^{EE} + \gamma C^{TE} \vphantom{\frac{1}{2}}
\right] \,\,,
\eea
and 
\bea
{\rm Var}(r) = \frac{1}{2} \sum_{l} (2l+1)
\left[  \vphantom{\frac{1}{2}} \alpha^2 (C^{TT})^{2} +
\beta^{2} (C^{EE})^{2} + \right. \nonumber \\
\frac{\gamma^{2}}{2} [ (C^{TE})^{2} + C^{EE}C^{TT}]  + 2\alpha\beta (C^{TE})^{2} + \nonumber \\
\left. \vphantom{\frac{1}{2}} + 2\beta\gamma C^{TE}C^{EE} + 2\alpha\gamma C^{TE}C^{TT} 
\right]\,\, \approx 2\langle r \rangle \,\,,
\eea
where we use standard contraction formulae to compute the four point functions, and the last
approximation is good when ${\bf C}$ is very close to ${\bf C}'$. The central limit theorem ensures that the
distribution of $r$ is well approximated by a Gaussian; we therefore obtain,
\be
P(r<0) = \frac{1}{2} \left[ 1 - {\rm erf}\left( \frac{\langle r \rangle}{\sqrt{2 {\rm Var}(r)}}
\right) \right]\,\,.
\ee

\begin{table}
\begin{tabular}{cccc}
\hline
Experiment & Beam & $10^{6} \Delta T/T$ & $10^{6} \Delta T/T$ \\
& FWHM (arcmin) & (I) & (Q,U) \\
\hline
\WMAP (V band) & 21 & 11.0 & 15.6 \\
\PLANCK (143 Ghz) & 7.1 & 2.2 & 4.2 \\
HiRes POL & 3.0 & 1.0 & 1.0 \\
Cosmic Variance & 0.0 & 0.0 & 0.0 \\
\hline
\end{tabular}
\caption{\label{tab:expspec} Detector sensitivities and beams for different CMB temperature and
polarization experiments. HiRes POL refers to a hypothetical all sky CMB polarization experiment.}
\end{table}

The above expressions can be generalized to take into account of detector noise; one
substitutes $C^{x} \rightarrow C^{x}+N^{x}$, where $x = TT,EE$ assuming
the temperature and polarization measurements are uncorrelated. The noise is determined
by the angular size of the telescope beam, and the temperature sensitivity of the detectors 
\cite{1995PhRvD..52.4307K, 2000ApJ...530..133T},
\be
N(l) = (w_{p})^{-1} \exp\left[l(l+1) \theta^{2}\right] \,\,,
\ee
where $\theta$ is related to the FWHM of the beam by ${\rm FWHM} = \theta \sqrt{8 \ln 2}$,
$(w_{p})^{-1/2} = \Delta T \times  {\rm FWHM}$, and all angles are in radians.

We now state our algorithm for the detectability of DM annihilation:
\begin{enumerate}
\item Consider two cosmological models, one with $\edm=0$ and the other with $\edm>0$. 
For simplicity, we assume a minimal cosmological model with 6 parameters ($\Omega_{M},
~\Omega_{b},~h,~n_{s},~A,~\tau$).
\item We adjust these parameters for the model with no DM annihilation to minimize $\langle r \rangle$.
\item At the minimum, we compute ${\rm Var}(r)$ and $P(r<0)$, thereby obtaining a measure of the 
detectability of $\edm$. 
\end{enumerate}
This algorithm is complementary to the Fisher information methods currently 
popular in cosmology. The Fisher information probes the likelihood function in the 
neighborhood of a fiducial point, estimating the minimal theoretically achievable
errors. On the other hand, we explicitly track the degeneracy
locus well beyond the immediate neighborhood of a fiducial model.

\subsection{Results}

\begin{figure}
\includegraphics[width=3in]{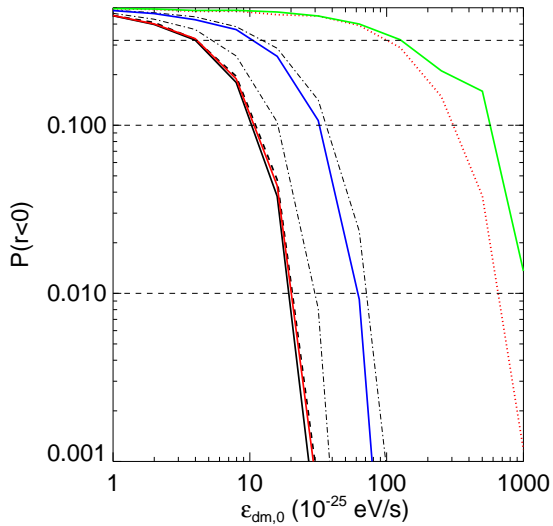}
\caption{\label{fig:degen} The detectability of a given injection energy, $\edmnow$, 
for the different experimental specifications of Table~\ref{tab:expspec}, 
assuming a maximum multipole, $l_{max}=2500$. The
lines, from left to right, are for cosmic variance, HiRes POL, the \PLANCK 143 Ghz
channel, and the \WMAP V band. The dashed line shows the cosmic variance detectability
for $l_{max}=1500$, while the dot-dashed lines show it 
for 20\% and 50\% sky coverage. The dotted line uses the parameters for HiRes POL,
but with no polarization information. Also shown are the 68\%, 90\%, and 99\% levels.}
\end{figure}

We consider four different experiments - a cosmic variance limited experiment, a hypothetical
high resolution polarization experiment (hereafter HiRes POL), the \PLANCK 143 Ghz channel 
\footnote{http://www.esa.int/science/planck}
and the \WMAP V band \cite{2000ApJ...530..133T}\footnote{http://lambda.gsfc.nasa.gov}. 
The parameters chosen for HiRes POL are similar to those projected for (eg. NASA's 
CMBpol \footnote{Now designated as the ``Inflation Probe''.}) 
missions designed to detect the imprint of gravitational 
radiation from inflation.
The assumed noise and beam characteristics
of these experiments are in Table~\ref{tab:expspec}, and the noise
power spectra are plotted in Fig.~\ref{fig:models}. Note that our aim here is to 
consider representative parameters, and not an optimization of experimental 
specifications.

\begin{table}
\begin{tabular}{cccc}
\hline
Experiment & & $\edmnow (10^{-25} \eV/{\rm s})$ & \\
& $68\%$ & $90\%$ & $99\%$ \\
\hline
\WMAP (V band) & 1.5e2 & 5.6e2 & 1.1e3 \\
\PLANCK (143 Ghz) & 1.1e1 & 3.2e1 & 6.2e1 \\
HiRes POL & 4.1 & 1.2e1 & 2.2e1 \\
Cosmic Variance & 4.0 & 1.1e1 & 2.1e1 \\
\hline
\end{tabular}
\caption{\label{tab:expsigma} Detectable values of $\edm$ at  68\%, 90\%, 
and 99\% confidence levels, for different experimental parameters (Table~\ref{tab:expspec}).
}
\end{table}

The results are summarized in Fig.~\ref{fig:degen} and Table~\ref{tab:expsigma}.
A principal feature is the significant improvement of \PLANCK
over \WMAP, due to the introduction of high S/N polarization data, 
evident in Fig.~\ref{fig:models}. The importance of polarization is further
emphasized by comparing HiRes POL, to an experiment with the same temperature
but no polarization sensitivity.
As discussed earlier, the temperature 
power spectrum suffers from a degeneracy between $n_{s}$ and $\edm$, and
therefore cannot constrain $\edm$ by itself. Introducing polarization breaks this 
degeneracy, allowing $\edm$ to be measured with significantly greater sensitivity.
We also observe that HiRes POL (almost) achieves 
the cosmic variance sensitivity limits.

\begin{figure}
\includegraphics[width=3in]{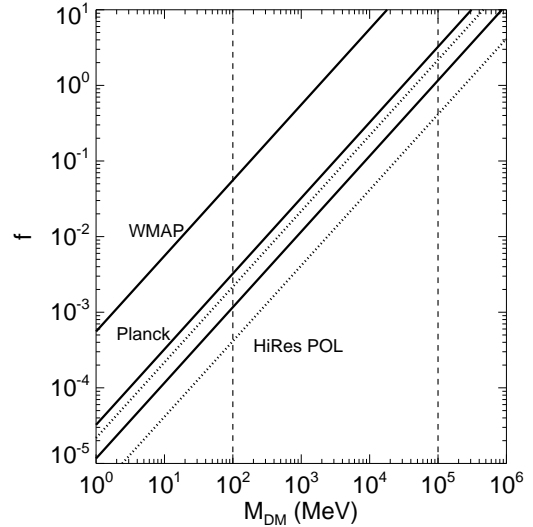}
\caption{\label{fig:exclude} The 90\% exclusion region in the $M_{DM} ~-~ f$ plane, for 
(from top to bottom) \WMAP, \PLANCK and HiRes POL; regions
above the lines can be excluded by these experiments. This assumes our fiducial cosmology, and 
Eq.~\ref{eq:dmtoedm2} to relate $f$ and $M_{DM}$ to $\edm$; $\sv$ is set to the value required 
for a thermal relic density, but deviations from this can be absorbed into $f$. The dotted lines
also show the 68\% and 99\% exclusion regions for our hypothetical polarization experiment.}
\end{figure}

The advantage of phrasing the limits 
in terms of $\edmnow$ is that they are independent of a particular DM model, or indeed, of any
mechanism for the injection of additional energy during recombination. However, 
for a single species of DM, 
Eq.~\ref{eq:dmtoedm} relates $M_{DM}$, $\Omega_{DM}$, $\sv$ and 
$f$ to $\edmnow$. Furthermore, $f$ and $\sv$ are degenerate with each other,
allowing us to translate 
our limits on $\edm$ into constraints in the $f$-$M_{DM}$ plane. Fig.~\ref{fig:exclude} does this 
for our fiducial cosmological model, using Eq.~\ref{eq:dmtoedm2} and Table~\ref{tab:expsigma}.
We observe that, assuming $f \sim 0.1 - 1$, \PLANCK will be able to detect (at 90\% confidence) DM
annihilation from particles with masses less than $\sim 3 ~-~ 30 \GeV$,
while HiRes POL increases that lower bound to $\sim 10~-~100 \GeV$.
These limits assume $\sv = 2 \times 10^{-26} \rm{~cm}^{3}/\rm{s}$, appropriate for a thermal relic;
a higher $\sv$ (due to eg. co-annihilations) would proportionally increase the limits.
We note that these limits probe relevant parts of parameter space,
and are complementary to accelerator and direct detections, since
they make very different assumptions.

It is timely to ask what parts of DM model space can be constrained by \WMAP. Fig.~\ref{fig:exclude}
shows that \WMAP will have no sensitivity to DM models with masses $> 1 \GeV$, and therefore, 
to the traditional DM candidates. However, \cite{2004NuPhB.683..219B,2004PhRvD..69j1302B,2004PhRvL..92j1301B}
have proposed a light ($\sim 10~-~100 \MeV$)
DM particle to explain the $511 \keV$ flux observed by the INTEGRAL satellite. Since the
particle mass is below the photon transparency window, 
the annihilation energy is efficiently converted into ionizations, suggesting that such models will be 
strongly constrained by the \WMAP polarization measurements.

\begin{figure}
\includegraphics[width=3in]{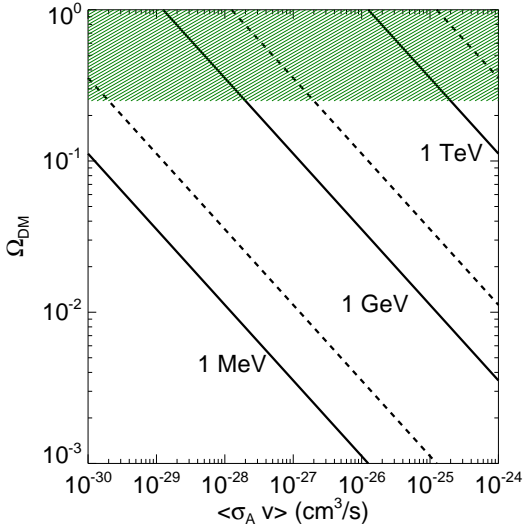}
\caption{\label{fig:omegasv} Contours in the $\Omega_{DM}$-$\sv$ plane 
with $\edmnow = 10^{-24} \eV/{\rm s}$, for $M_{DM}/f$ = 1 TeV, 1 GeV, and
1 MeV (solid), and 10 TeV, 10 GeV, and 10 MeV (dashed). Regions above these
contours are accessible to an experiment with the sensitivity to measure $\edmnow 
= 10^{-24} \eV/{\rm s}$. The shaded region shows the region excluded by our fiducial
model.}
\end{figure}

The discussion above focuses on the detection of an unknown particle, and 
we therefore have concentrated on the simplest case of a single species of DM. Reality
could well be more complicated, and the DM could well consist of multiple species.
In such a case, in addition to determining the nature of the DM, we would need to 
determine the relative contribution of the different species. Given a particle 
with a known mass and annihilation cross section, constraints on $\edm$ put an upper bound
on the density of these particles, without making any assumptions about 
whether the particle is a thermal relic or not. Fig.~\ref{fig:omegasv} demonstrates this
for different values of $M_{dm}$, $\sv$ and $f$.
This emphasizes the complementary nature of astrophysical and particle physics probes of
DM; at low masses, the CMB is competitive with direct detection and accelerator searches
of DM. However, given a detected DM particle, the CMB probes, with minimal 
assumptions, its cosmological density, that in turn, has the potential to 
constrain theoretical models for its formation.

\section{Other Astrophysical Constraints}
\label{sec:other}

Does DM annihilation during the recombination epoch have other 
observational consequences? We consider three possibilities - distortions
to the CMB spectrum, redshifted photons contributing to the diffuse photon
background, and molecular hydrogen production.

We start by considering the additional energy density injected per 
unit time at $z=1000$ due to DM annihilation,
\bea
{\cal E} & &= \sv M_{DM} n_{DM}^2 \nonumber \\
& &\sim 2\times 10^{-13} \eV/{\rm cm}^{3}/{\rm s} \left(\frac{1+z}{1000}\right)^{6} \times \nonumber \\
& & \left(\frac{\Omega_{DM} h^{2}}{0.12}\right)^{2} 
\left(\frac{\sv}{2 \times 10^{-26} \cm^{3}/{\rm s}} \frac{100 \GeV}{M_{DM}}\right) \,\,.
\eea
Over a Hubble time, the energy injected is 
\bea
E & &\sim 2 \eV/{\rm cm}^{3} \left(\frac{1+z}{1000}\right)^{9/2}  \left(\frac{0.147}{\Omega_{M}h^{2}}\right)^{1/2} 
\times \nonumber \\
& & \left(\frac{\Omega_{DM} h^{2}}{0.12}\right)^{2}
\left(\frac{\sv}{2 \times 10^{-26} \cm^{3}/{\rm s}} \frac{100 \GeV}{M_{DM}}\right) \,\,.
\label{eq:dmphotonenergy}
\eea
This is much less than the energy density of the CMB,
\be
E_{CMB} \sim 0.25 \times 10^{12} \left(\frac{1+z}{1000}\right)^{4} \eV/{\rm cm}^{3} \,\,,
\ee
implying that the distortions are below detection thresholds. We also note that although during
matter domination, the energy from DM annihilation is growing faster than the CMB energy density, 
the exponent of $9/2$ changes to $4$ during radiation domination, and so the injected energy density
from DM annihilation never is a substantial fraction of the CMB energy density.

Could photons from DM annihilations, injected into an optically thin 
part of the spectrum, be detected as part of the diffuse photon background today? 
The transparency region from $10^{8}$ to $10^{10} \eV$ in Fig.~\ref{fig:photonlosses}
implies that these photons would have energies between $10^{5}$ and $10^{7} \eV$ today.
Furthermore, Eq.~\ref{eq:dmphotonenergy} for standard parameters gives a present-day photon energy 
density of $E \sim 2 \times 10^{-12} \eV/{\rm cm}^{3}$, implying a flux of 
$0.5 \times 10^{-2} \eV/{\rm cm}^{2}{\rm /s/sr}$. Comparing this to the 
observed flux of $1 \times 10^{3} \eV/{\rm cm}^{2}{\rm /s/sr}$ 
\cite{1998ApJ...494..523S,2000AIPC..510..467W,2004ApJ...613..956S}, we find 
that the fraction possibly due to DM annihilation is considerably below
the uncertainties in the measurement.

The residual ionization after recombination serves as a catalyst for the 
production of molecular hydrogen, first via $2H + H^{+} \rightarrow H_{2} + H^{+}$
at $z \sim 500$, and then by $2H + e^{-} \rightarrow H_{2} + e^{-}$ at $z \sim 100$.
Molecular hydrogen is important since it serves as a coolant, allowing for the 
collapse of the first cosmological objects. A higher ionization fraction could, 
in principle, create a greater abundance of molecular hydrogen, triggering 
collapse at an earlier epoch.
The equation for the evolution of the 
molecular hydrogen fraction, $f \equiv n[H_{2}]/n[H]$, is \cite{1997ApJ...474....1T}
\be
H(a) \frac{df}{d \ln a} = k_{eff} (1 - x - 2f) n x  \,\,,
\label{eq:molh2}
\ee
where $x$ is the ionization fraction and $n = n[H] + n[H^{+}] + n[H_{2}]$ is
the density of hydrogen atoms, and $k_{eff}$ is the effective rate coefficient,
approximately constant for $x,f \ll  1$. For $x \ll 1$, 
Eq.~\ref{eq:molh2} describes a catalyst-starved reaction; increasing $x$
proportionally increases the $H_{2}$ fraction. The standard 
recombination scenario produces an $H_{2}$ fraction of $\sim 10^{-6}$ 
\cite{1998A&A...335..403G}
However, the fraction required to efficiently cool halos is approximately
$10^{-3}$; primordial $H_{2}$, even with an 
enhanced ionization fraction, is too small to significantly alter 
structure formation. The required $H_{2}$ is 
produced in regions of high density, where the reaction rates are significantly 
higher. Unfortunately, this process is relatively insensitive to initial conditions
\cite{1997ApJ...474....1T},
making it difficult to constrain the $H_{2}$ and ionization fractions.

\section{Discussion}
\label{sec:discuss}

We have considered the effect that DM annihilation has on the recombination history
of the universe, and therefore, on the CMB temperature and polarization power spectra.
We argued that the effect of DM annihilation on the IGM is well approximated 
by an injection of a fraction $f$ of the rest mass energy of the DM particles into the IGM, 
where it is instantaneously used to heat and ionize the IGM. This ``on the spot'' approximation
allowed us to compute the altered recombination history; the epoch of recombination is
unchanged, but the residual ionization fraction increases. This broadens the visibility
function, suppressing the temperature power spectrum, but enhancing the polarization power
spectrum. Furthermore, the thicker visibility function shifts the peaks in TE and EE power spectra
relative to the peaks in the TT power spectrum.

Given the modified power spectra, we can ask whether the changes are detectable, or if they are 
degenerate with other cosmological parameters? A cosmic variance limited survey
is sensitive to an energy injection rate of $\edmnow \sim 10^{-24} \eV/{\rm s}$, probing 
masses $\la 100 \GeV$.
Furthermore, these limits are attainable by CMB experiments designed to detect the polarization
created by the stochastic gravitational wave background. The limits for \PLANCK are about an
order of magnitude worse, and the limits are significantly degraded for \WMAP, due to its 
polarization sensitivity. However, \WMAP will be able to constrain low mass (${\cal O}(10 \MeV
~-~ 1 \GeV)$) DM particles such as those proposed by \cite{2004PhRvL..92j1301B}.

We have kept our analysis as generic and idealized as possible, to ensure that
our results are independent of the particulars of any DM model. We now consider 
some of the issues ignored by the analysis above.

\begin{itemize}
\item\emph{How does one calculate $f$?} In order for a DM model to make a falsifiable prediction,
it is important to be able to calculate $f$ for a given model.
A simple algorithm to compute $f$ given all the decay channels and their branching 
ratios (admittedly a tedious task!) is :
\begin{itemize}
\item Compute the $e^{\pm}$ and photon energy spectra resulting from an annihilation.
\item Evolve the photon and $e^{\pm}$ spectra with the processes in Fig.~\ref{fig:cascade}.
\item Redshift the spectra to the next time and repeat.
\end{itemize}
One can obtain a qualitative picture of the results by considering Fig.~\ref{fig:photonlosses}. For 
photons below $\sim 10^{8} \eV$ and electrons below $10^{10} \eV$, the energy is  efficiently deposited
($\sim 0.1~-~1$) into the IGM. At higher energies, the efficiency is dictated by efficiency of pair production; however,
even in this case, one would expect a non-negligible ($\sim 0.01-0.1$) fraction 
to be converted into ionizations and heat. One consequence is that one would expect $f$ to generically
increase with decreasing mass, weakening our ability to constrain the highest masses but 
strengthening constraints on lower masses. Finally, to compare to plots such as 
Fig.~\ref{fig:exclude}, one must also include variations in $\sv$ in $f$. While $\sv$
is only logarithmically dependent on $M_{DM}$ for a thermal
relic density of a single species, it can vary significantly if one includes the possibilities
of multiple DM species or co-annihilations. Just how generic these processes are will depend
on the class of theories being considered; initial attempts to answer such questions 
within the framework of the MSSM are in \cite{2004JHEP...10..052B}.

\item \emph{Is the standard recombination calculation accurate enough?} Detecting DM annihilation 
by its effect on the recombination history requires that we understand the fiducial
recombination physics to better than 1\%. The current recombination calculation 
is accurate to $\sim 1\%$ \cite{2000ApJS..128..407S}, although there recently
have been claims of corrections due to two photon processes\cite{2005astro.ph..1672D}. Although the current 
calculation may not be accurate enough, we do not know of any theoretical limit that would
prevent reaching the required accuracy. We should also emphasize that this accuracy is 
only necessary for detecting injection energies $\sim 10^{-25} \eV/{\rm s}$; the current
calculation is accurate enough to detect higher energies.

\item \emph{What about real-world complications?} The limits on DM annihilation projected
in Sec.~\ref{sec:detect} assumed an idealized CMB experiment with full sky coverage,
and no contamination to the primary CMB due to Galactic and extragalactic foregrounds.
We briefly discuss these below; we however emphasize that these are not complications
peculiar to measuring DM annihilation, but will affect any high precision polarization
experiment (eg. detecting the gravitational wave background in the CMB).

Decreasing the sky coverage reduces the number of available modes, increasing 
the errors due to cosmic variance and degrading our ability to distinguish between
models. A full sky CMB experiment at these sensitivities might survey an effective
sky fraction of 50\% due to Galactic and point source cuts, increasing the errors
by $\sqrt{2}$; the impact of which is shown in Fig.~\ref{fig:degen}.
Even after excluding the most contaminated regions of the mask, it will be 
necessary to separate Galactic and extragalactic foregrounds using
their frequency dependence. This problem has been examined in detail
by a number of authors \cite[eg.][]{2000ApJ...530..133T}, who demonstrate that this separation is
possible on large angular scales. Furthermore, extragalactic foregrounds
start to dominate only on small angular scales ($l \sim 2500$), while the 
signal from DM annihilation is maximal on larger scales $l < 1500$ (Fig.~\ref{fig:degen}), 
suggesting that foregrounds are a tractable problem.
\end{itemize}

We conclude by reiterating the complementarity between the different probes of DM. 
As seen above, the CMB is able to put strong constraints on the cosmological
abundance of a light ($\ll 100 \GeV$) DM particle that evades accelerator 
searches. At higher masses ($> 100 \GeV$), the injected energy from annihilations 
decreases and the CMB is no longer competitive with other searches. However, given
a DM candidate with a known mass and annihilation cross section, the
injection energy constraints from the CMB translate into a constraint on the
density of these particles. Importantly, this does not assume that the 
particle is a thermal relic, allowing us to constrain the particle content
of the DM, including particles that produced via non-equilibrium processes.

Understanding the properties of the dark matter remains one of the 
most important problems in cosmology today. And astrophysical probes, such as 
the one discussed here, have an important role to play, both in detecting
possible candidates and understanding their cosmological abundance.

\acknowledgments
Our approach to this problem was crucially influenced by long
conversations with Latham Boyle, Jim Peebles, Uros Seljak and David Spergel.
We also thank Bruce Draine, Jim Gunn, Chris Hirata, Kevin Huffenberger, and
Lyman Page for their encouragement, criticisms, and suggestions. 
DPF is supported by NASA LTSA grant
NAG5-12972, while NP acknowledges the New York Public Library
where sections of this paper were written.
This research made use of the NASA
Astrophysics Data System (ADS) and the IDL Astronomy User's Library at
Goddard\footnote{http://idlastro.gsfc.nasa.gov/}. 

\bibliography{biblio,preprints}

\end{document}